\documentclass[aps,prd,amsmath,twocolumn,amssymbaps,showpacs]{revtex4-1}
\usepackage{graphicx}  
\usepackage{dcolumn}   
\usepackage{bm}        
\usepackage{amssymb}   
\usepackage{amsmath}   
\usepackage{bbm}
\usepackage{draftcopy}
\usepackage{relsize} 
\usepackage{xfrac}    
\usepackage{slashed}

\usepackage{color}
\definecolor{dgreen}{cmyk}{1.,0.,1.,0.2}        
\definecolor{orange}{cmyk}{0.,0.353,1.,0.}    

\usepackage[bookmarks]{hyperref}

\newcommand{\di}{{\rm d}}

\newcommand{\tr}{{\rm tr}}

\newcommand{\be}{\begin{equation}}
\newcommand{\ee}{\end{equation}}                                                                               
\newcommand{\bea}{\begin{eqnarray}}
\newcommand{\eea}{\end{eqnarray}}

\begin{document}
\title{Reentrant pion superfluidity and cosmic trajectories within PNJL model}
\author{Gaoqing Cao$^1$} \author{Lianyi He$^2$} \author{Pengming Zhang$^1$}
	\address{$^1$School of Physics and Astronomy, Sun Yat-sen University, Zhuhai 519088, China\\$^2$Department of Physics, Tsinghua University, Beijing 100084, China}

\date{\today}

\begin{abstract}
In this work, we self-consistently explore the possibility of charged pion superfluidity and cosmic trajectories in early Universe under the framework of Polyakov-Nambu--Jona-Lasinio model. By taking the badly constrained lepton flavor asymmetries $l_{\rm e}$ and $l_\mu$ as free parameters, the upper boundaries of pion superfluidity phase are consistently found to be around the pseudocritical temperature at zero chemical potentials. So the results greatly support the choice of $T=0.16~{\rm GeV}$ as the upper boundary of pion superfluidity in the previous lattice QCD study. Take $l_{\rm e}+l_\mu=-0.2$ as an example, we demonstrate the features of pion condensation and the associated cosmic trajectories with the evolution of early Universe. While the trajectory of electric chemical potential reacts strongly at both the lower and upper boundaries of reentrant pion superfluidity, the trajectories of other chemical potentials only respond strongly at the upper boundary.
\end{abstract}


\maketitle

\section{Introduction}

As we know, the field of high energy nuclear physics (HENP) initiated with the search of quark-gluon plasma (QGP) phase~\cite{Shuryak:1980tp} for quantum chromodynamics (QCD) matter, and the QGP phase was expected to be realized through relativistic heavy ion collisions (HICs)~\cite{Lee:1982qw,Heinz:2000bk}. Up to now, the existence of QGP at high temperature becomes a consensus in HENP -- a direct evidence is the observation of number-of-constituent quark scaling of the elliptic flows for mesons and baryons in large center of mass HICs~\cite{Adams:2005zg}. Of course, other properties of QGP have also been well explored and one remarkable discovery is that the QGP is a nearly perfect liquid~\cite{Kovtun:2004de,Kolb2004}. Nevertheless, a much more sophisticated and challenging mission is to depict the first QCD phase diagram in $T-\mu_{\rm B}$ plane with the help of HICs. At early time, people didn't find it a real phase transition from QGP to hadron phase at small baryon number density $n_{\rm B}$ according to either lattice QCD simulations~\cite{Aoki:2006we,Bhattacharya:2014ara} or experimental detections~\cite{Floris:2014pta,Adamczyk:2017iwn}. Recently, the STAR group is carrying out lower energy collisions in their BES II experiments with the hope of catching the critical end point by increasing $n_{\rm B}$~\cite{Luo:2017faz}.

Besides, the $T-\mu_{\rm I}$ phase diagram has also been extensively studied and charged pion ($\pi^\pm$) superfluidity was expected theoretically in the region with not too large $T$ and large isospin chemical potential $\mu_{\rm I}~(>m_\pi^{\rm v})$~\cite{Son:2000xc,Kogut:2002zg,Brandt:2017oyy,He:2005nk}. Specifically, the transition between chiral symmetry breaking or restoration phase and $\pi^\pm$ superfluidity was consistently found to be of second-order in chiral perturbation theory~\cite{Son:2000xc}, lattice QCD~\cite{Kogut:2002zg,Brandt:2017oyy} and effective models such as Nambu--Jona-Lasinio (NJL) model~\cite{He:2005nk}. And with the increasing of $\mu_{\rm I}$, the Bose-Einstein condensation of $\pi^\pm$ was found to smoothly crossover to the Bardeen-Cooper-Schieffer phase~\cite{Sun:2007fc} which then possibly becomes the quarksonic matter~\cite{Cao:2016ats}. However, it's a pity that the systems in nature with large $\mu_{\rm I}$, neutron stars, are also large $\mu_{\rm B}$ matters, and the constraint of electric neutrality eventually disfavors $\pi^\pm$ superfluidity in cold neutron stars~\cite{Andersen:2007qv,Abuki:2008tx}. 

The hope of finding $\pi^\pm$ superfluidity in nature reburned in the explorations of proto-neutron stars~\cite{Abuki:2009hx} and early Universe~\cite{Middeldorf-Wygas:2020glx,Vovchenko:2020crk}, where the temperature and lepton flavor densities can be much larger. Especially, large lepton flavor densities would also help to stabilize $\pi^\pm$ mesons against weak decays thanks to the Pauli blocking effect to the final state~\cite{Abuki:2009hx}. And it is interesting that the QCD phase would impact primordial gravitational wave (GW) and  generation rate of black holes quite well in early Universe~\cite{Vovchenko:2020crk}. According to the Big Bang theory, temperature drops to $\sim200\,{\rm MeV}$ in $10^{-6}\,{\rm s}$ of the Big Bang and the early Universe enters the QCD epoch where strong interaction dominates particle scattering. In the QCD epoch, the baryon and lepton number $U(1)$ symmetries were already violated, and the tracing back of the observations in present Universe constrains the corresponding number densities as $n_{\rm B}/s=8.6*10^{-11}$~\cite{Ade2016} and $|n_{\rm l}/s|<0.012$~\cite{Oldengott2017} with $s$ the entropy density. Under these and electric neutrality constraints, the cosmic trajectories of several chemical potentials were explored by combining hadron resonance gas model, lattice QCD and free quark gas model~\cite{Wygas:2018otj}. Later, the possibility of $\pi^\pm$ superfluidity was realized at large lepton flavor asymmetries~\cite{Middeldorf-Wygas:2020glx,Vovchenko:2020crk} and the lower phase boundary was reasonably depicted by utilizing the criteria $|\mu_{\rm Q}|>m_{\pi^\pm}(\mu_{\rm Q},T)$~\cite{Vovchenko:2020crk}, where $\mu_{\rm Q}$ plays the role of $\mu_{\rm I}$ for charged pions.

In Ref.~\cite{Vovchenko:2020crk}, the temperature effect on $m_{\pi^\pm}(\mu_{\rm Q},T)$, the order parameter of their effective mass model, was only taken into account through the ideal gas part. This might be the reason why the lattice QCD inspired model couldn't predict the upper phase boundary. And it is also a pity that they evaluated the phase boundary without demonstrating the evolutions of the true order parameter: charged pion condensates. The advantage of NJL model is that chiral symmetry breaking and restoration, pion superfluidity, and the corresponding pion masses can be self-consistently studied by solving the gap equations and the zero points of pion propagators, all of which can be derived analytically, see Ref.~\cite{He:2005nk}. So we intend to recheck the phase boundary of $\pi^\pm$ superfluidity in the Polyakov loop extended NJL model (PNJL model)~\cite{Fukushima:2017csk} and show how the order parameters and cosmic trajectories evolve across the $\pi^\pm$ superfluidity phases. In principle, the PNJL model  is able to mimic QCD more realistically by counting the deconfinement effect to quarks and gluon contributions to the total entropy~\cite{Fukushima:2017csk}.

The paper is organized as follows. In Sec.\ref{formalism}, we develop the overall formalism for the study of the QCD and quantum electroweak dynamics (QEWD) matter in the QCD epoch. For the QCD sector, the two- and three-flavor PNJL models will be laid out explicitly in Sec.\ref{formalism2f} and Sec.\ref{formalism3f}, respectively. The 
QEWD sector will be approximated as free gases in Sec.\ref{QEWD}. Then, we present numerical results in Sec.\ref{numerical} and finally summarize in Sec.\ref{summary}.

\section{The overall formalism}\label{formalism}

In the QCD epoch of early Universe, both QCD and QEWD sectors are relevant: the elementary degrees of freedom are quarks and gluons in the QCD sector and leptons and photons in the QEWD sector. To study the QCD sector self-consistently, we adopt the chiral effective Polyakov-Nambu--Jona-Lasinio (PNJL) model, where quarks contribute through the NJL model part and the contributions of gluons are given in terms of Polyakov loop (PL) according to the lattice QCD simulations~\cite{Meisinger:1995ih,Fukushima:2003fw,Ratti:2005jh,Fukushima:2017csk}. For the QEWD sector, the coupling constants are usually small, so we are satisfied to utilize free gas approximation for the involved particles. The following sections are devoted to developing detailed formalisms for both the QCD matter with two or three flavors and the free QEWD matter.

\subsection{The QCD sector with two flavors}\label{formalism2f}
The Lagrangian density of two-flavor PNJL model with electric charge chemical potential $\mu_{\rm Q}$ and baryon chemical potential $\mu_{\rm B}$ can be given as~\cite{Fukushima:2017csk,Klevansky:1992qe,Hatsuda:1994pi}
\begin{eqnarray}
{\cal L}&=&\bar\psi\left[i\slashed{\partial}-i\gamma^4\left(ig{\cal A}^4+Q_{\rm q}\mu_{\rm Q}+{\mu_{\rm B}\over3}\right)-m_0\right]\psi\nonumber\\
&&+G\left[\left(\bar\psi\psi\right)^2+\left(\bar\psi i\gamma_5\boldsymbol\tau\psi\right)^2\right]-V(L,L^*)
\end{eqnarray}
in Euclidean space. In the NJL model part, $\psi=(u,d)^T$ represents the two-flavor quark field and ${\cal A}^4=A^{\rm 4c}T^{\rm c}/2$ is the non-Abelian gauge field with $T^{\rm c}$ the Gell-Mann matrices in color space; $m_0\equiv m_{\rm 0l}\mathbbm{1}_2$ is the current mass matrix, the charge number matrix is
\bea
 Q_{\rm q}\equiv{\rm diag}(q_{\rm u},q_{\rm d})={1\over3}{\rm diag}(2,-1),
 \eea and ${\boldsymbol{\tau}}$ are Pauli matrices in flavor space. The pure gluon potential is given as a function of the Polyakov loop $$L={1\over N_{\rm c}}\tr\,e^{ig\int\di x_4{\cal A}^4}$$ and its complex conjugate $L^*$ by fitting to the lattice QCD data, that is,
 \bea
{V(L,L^*)\over T^4}&=&-{1\over2}\left(3.51-{2.47\over\tilde{T}}+{15.2\over\tilde{T}^{2}}\right)|L|^2-{1.75\over\tilde{T}^{3}}\nonumber\\
&&\times\ln\left[1-6|L|^2+4(L^3+{L^*}^3)-3|L|^4\right]
\eea
with $\tilde{T}\equiv T/T_0$ and $T_0=0.27\,{\rm GeV}$~\cite{Fukushima:2017csk}.

To obtain the analytic form of the basic thermodynamic potential, we take Hubbard-Stratonovich transformation with the help of the auxiliary fields $\sigma=-2{G}\bar{\psi}\psi$ and ${\boldsymbol{\pi}}=-2{G}\bar{\psi}i\gamma^5{\boldsymbol{\tau}}\psi$~\cite{Klevansky:1992qe} and the Lagrangian becomes
\begin{eqnarray}
{\cal L}&=&\bar{\psi}\!\left[i{\slashed \partial}\!-\!i\gamma^4\!\left(\! ig{\cal A}^4\!+\!Q_{\rm q}\mu_{\rm Q}\!+\!{\mu_{\rm B}\over3}\!\right)\!-\!i\gamma^5\boldsymbol{\tau}\cdot\boldsymbol{\pi}\!-\!\sigma\!-\!m_0\right]\!\psi\nonumber\\
&&-{\sigma^2+\boldsymbol{\pi}^2\over4G}-V(L,L^*).
\end{eqnarray}
For later convenience, we alternatively represent it as
\bea
{\cal L}\!&=&\!\bar{\psi}\!\Big[i{\slashed \partial}\!-\!i\gamma^4\!\!\left(ig{\cal A}^4+Q_{\rm q}\mu_{\rm Q}\!+\!{\mu_{\rm B}\over3}\right)\!\!-\!i\gamma^5\!\!\left({\tau_3}{\pi^0}\!+\!{\tau_{\pm}}{\pi^\pm}\right)\nonumber\\
&&\!-\sigma-m_0\Big]\!\psi-{\sigma^2+{\left({\pi^0}\right)^2}+2{\pi^+}{\pi^-}\over4G}-V(L,L^*)
\eea
in the forms of physical particles: $\pi^0=\pi_3$ and $\pi^\pm={1\over\sqrt{2}}(\pi_1\mp i\pi_2)$, where $\tau_\mp={1\over\sqrt{2}}(\tau_1\mp i\tau_2)$ is the lowering/raising operator in flavor space. To our present interests, it is enough to assume the expectation values of the auxiliary fields to be $$\langle\sigma\rangle=m-m_0,\ \langle\pi^0\rangle=0,\ \langle\pi^\pm\rangle=\Pi/\sqrt{2}.$$ Then, in mean field approximation, the thermodynamic potential can be given in energy-momentum space as
\begin{eqnarray}
\Omega_{\rm 2f}&=&-{\rm Tr}\ln\left[{\slashed k}\!-\!m\!-\!i\gamma^4\!\!\left(ig{\cal A}^4\!+\!Q_{\rm q}\mu_{\rm Q}\!+\!{\mu_{\rm B}\over3}\right)\!-\!i\gamma^5\tau_{1}\Pi\right]\nonumber\\
&&+{(m-m_0)^2+\Pi^2\over4G}+V(L,L^*)
\end{eqnarray}
with the trace ${\rm Tr}$ over the energy-momentum, spinor, flavor, and color spaces. To derive the explicit form of the trace term, we need to solve the quark dispersions 
from the zero points of their inverse propagator in Minkowski space, that is, from
\begin{eqnarray}
{\rm Det}\!\left[{\slashed k}\!-\!m\!+\!\gamma^0\!\!\left(ig{\cal A}^4\!+\!Q_{\rm q}\mu_{\rm Q}\!+\!{\mu_{\rm B}\over3}\right)\!-\!i\gamma^5\tau_{1}\Pi\right]=0.
\end{eqnarray}
We get $k_0=E^{\rm t}(k)\pm\left({\mu_{\rm Q}+2\mu_{\rm B}\over6}+i\langle g{\cal A}^4\rangle\right)$ with~\cite{He:2005nk} $$E^{\pm}(k)=\sqrt{\left(\epsilon(k)\pm{\mu_{\rm Q}\over2}\right)^2+\Pi^2},\ \epsilon(k)=\sqrt{k^2+m^2}.$$ Finally, in the saddle point approximation~\cite{Fukushima:2017csk}: $L=L^*$, the thermodynamic potential can be given directly as~\cite{Xiong:2009zz}
\begin{widetext}
\begin{eqnarray}
\Omega_{\rm 2f}&=&V(L,L)+{(m-m_0)^2+\Pi^2\over4G}-2N_c\int^\Lambda{\di^3k\over(2\pi)^3}\sum_{t=\pm}E^{\rm t}(k)-2T\!\!\int\!\!{\di^3k\over(2\pi)^3}\!\sum_{t,u=\pm}Fl\left(L,u,E^{\rm t}(k),{\mu_{\rm Q}\!+\!2\mu_{\rm B}\over6}\right),\\
&&Fl(L,u,x,y)=\log\left[1+3L\,e^{-{1\over T}\left(x-u\,y\right)}+3L\,e^{-{2\over T}\left(x-u\,y\right)}+e^{-{3\over T}\left(x-u\,y\right)}\right],\nonumber
\end{eqnarray}
 where three-momentum cutoff $\Lambda$ is adopted to regularize the divergent vacuum term.

Armed with the equation of state, the coupled gap equations follow the minimal conditions, $\partial_{\rm m}\Omega_{\rm 2f}=\partial_{\rm \Pi}\Omega_{\rm 2f}=\partial_{\rm L}\Omega_{\rm 2f}=0$, as
\bea
0&=&{m-m_0\over 2G}-2N_c\int^\Lambda{\di^3k\over(2\pi)^3}\sum_{t=\pm}{m\over\epsilon(k)}{\epsilon(k)\!+\!t{\mu_{\rm Q}\over2}\over E^{\rm t}(k)}+6\int{\di^3k\over(2\pi)^3}\sum_{t,u=\pm}{m\over\epsilon(k)}{\epsilon(k)\!+\!t{\mu_{\rm Q}\over2}\over E^{\rm t}(k)}dV_1\left(L,u,E^{\rm t}(k),{\mu_{\rm Q}\!+\!2\mu_{\rm B}\over6}\right),\label{gapl}\\
0&=&{\Pi\over 2G}-2N_c\int^\Lambda{\di^3k\over(2\pi)^3}\sum_{t=\pm}{\Pi\over E^{\rm t}(k)}+6\int{\di^3k\over(2\pi)^3}\sum_{t,u=\pm}{\Pi\over E^{\rm t}(k)}dV_1\left(L,u,E^{\rm t}(k),{\mu_{\rm Q}\!+\!2\mu_{\rm B}\over6}\right),\label{gappi}\\
0&=&T^4\left[-\left(3.51\!-{2.47\over\tilde{T}}\!+\!{15.2\over\tilde{T}^{2}}\right)L\!+\!{1.75\over\tilde{T}^{3}}{12L(1-L)^2\over1\!-\!6L^2\!+\!8L^3\!-\!3L^4}\right]-6T\!\int\!\!{\di^3k\over(2\pi)^3}\!\sum_{t,u=\pm}dV_2\left(L,u,E^{\rm t}(k),{\mu_{\rm Q}\!+\!2\mu_{\rm B}\over6}\right),
\eea
where we've defined two dimensionless auxiliary functions for future use:
\bea
dV_1(L,u,x,y)&=&{L\,e^{-{1\over T}\left(x-u\,y\right)}+2L\,e^{-{2\over T}\left(x-u\,y\right)}+e^{-{3\over T}\left(x-u\,y\right)}\over1+3L\,e^{-{1\over T}\left(x-u\,y\right)}+3L\,e^{-{2\over T}\left(x-u\,y\right)}+e^{-{3\over T}\left(x-u\,y\right)}},\\
dV_2(L,u,x,y)&=&{e^{-{1\over T}\left(x-u\,y\right)}+e^{-{2\over T}\left(x-u\,y\right)}\over1+3L\,e^{-{1\over T}\left(x-u\,y\right)}+3L\,e^{-{2\over T}\left(x-u\,y\right)}+e^{-{3\over T}\left(x-u\,y\right)}}.
\eea  
Furthermore, the entropy, electric charge number, and baryon number densities can also be derived analytically according to the thermodynamic relations as
\bea
\!\!\!\!\!\!s_{\rm 2f}&=&-{\partial \Omega_{\rm 2f}\over \partial T}=2\!\!\int\!\!{\di^3k\over(2\pi)^3}\!\sum_{t,u=\pm}\left[Fl\left(\!L,u,E^{\rm t}(k),{\mu_{\rm Q}\!+\!2\mu_{\rm B}\over6}\!\right)\!+\!{3\left(\!E^{\rm t}(k)\!-\!u\,{\mu_{\rm Q}\!+\!2\mu_{\rm B}\over6}\!\right)\over T}dV_1\left(\!L,u,E^{\rm t}(k),{\mu_{\rm Q}\!+\!2\mu_{\rm B}\over6}\!\right)\right]\nonumber\\
&&+T^3\left\{{1\over2}\left(4\times3.51-3\times{2.47\over\tilde{T}}+2\times{15.2\over\tilde{T}^{2}}\right)L^2+{1.75\over\tilde{T}^{3}}\ln\left[1-6L^2+8L^3-3L^4\right]\right\},\label{s2f}\\
\!\!\!\!\!\!n_{\rm Q}^{\rm2f}&=&-{\partial \Omega_{\rm 2f}\over \partial \mu_{\rm Q}}=N_c\!\!\int^\Lambda\!\!{\di^3k\over(2\pi)^3}\!\sum_{t=\pm}t{\epsilon(k)\!+\!t{\mu_{\rm Q}\over2}\over E^{\rm t}(k)}-\!3\!\!\int\!\!{\di^3k\over(2\pi)^3}\sum_{t,u=\pm}t{\epsilon(k)\!+\!t{\mu_{\rm Q}\over2}\over E^{\rm t}(k)}dV_1\left(\!L,u,E^{\rm t}(k),{\mu_{\rm Q}\!+\!2\mu_{\rm B}\over6}\!\right)\nonumber\\
&&\ \ \ \ \ \ \ \ \ \ \ \ \ +\int\!\!{\di^3k\over(2\pi)^3}\!\sum_{t,u=\pm}u\,dV_1\left(\!L,u,E^{\rm t}(k),{\mu_{\rm Q}\!+\!2\mu_{\rm B}\over6}\!\right),\label{nQ2f}\\
\!\!\!\!\!\!n_{\rm B}^{\rm2f}&=&-{\partial \Omega_{\rm 2f}\over \partial \mu_{\rm B}}=2\int{\di^3k\over(2\pi)^3}\!\sum_{t,u=\pm}u\,dV_1\left(\!L,u,E^{\rm t}(k),{\mu_{\rm Q}\!+\!2\mu_{\rm B}\over6}\!\right).\label{nB2f}
\eea
\end{widetext}

\subsection{The QCD sector with three flavors}\label{formalism3f}
In order to explore the properties of QCD matter more realistically, we adopt the three-flavor PNJL model where more low-lying mesons are involved and the QCD $U_{\rm A}(1)$ anomaly has been properly taken into account through the 't Hooft term. In saddle point approximation, the corresponding Lagrangian density can be given by~\cite{Fukushima:2017csk,Klevansky:1992qe,Hatsuda:1994pi}:
\begin{eqnarray}
{\cal L}_{\rm NJL}\!&=&\!-\!V(L,L)\!+\!\bar\psi\!\left[i\slashed{\partial}\!-\!i\gamma^4\!\!\left(\!ig{\cal A}^4\!+\!Q_{\rm q}\mu_{\rm Q}\!+\!{\mu_{\rm B}\over3}\!\right)\!-\!m_0\right]\!\psi\nonumber\\
&&+G\sum_{a=0}^8\left[(\bar\psi\lambda^a\psi)^2+(\bar\psi i\gamma_5\lambda^a\psi)^2\right]+{\cal L}_{\rm tH},
\end{eqnarray}
where $\psi=(u,d,s)^T$ is now the three-flavor quark field. Similar to the two-flavor case, the current mass and electric charge number matrices of quarks are respectively
 \bea
 m_0&\equiv&{\rm diag}(m_{\rm 0u},m_{\rm 0d},m_{\rm 0s}),\nonumber\\
 Q_{\rm q}&\equiv&{\rm diag}(q_{\rm u},q_{\rm d},q_{\rm s})={1\over3}{\rm diag}(2,-1,-1);
 \eea
the interaction index $\lambda^0=\sqrt{2/3}~\mathbbm{1}_3$ and $\lambda^i~(i=1,\dots,8)$ are Gell-Mann matrices in flavor space. For later use, the 't Hooft term ${\cal L}_{\rm tH}\equiv-K\sum_{t=\pm}{\rm Det}~\bar\psi\Gamma^t\psi$ can be represented as
\bea
\!\!\!\!\!\!\!\!{\cal L}_{\rm tH}\!=\!-{K\over2}\sum_{t=\pm}\epsilon_{ijk}\epsilon_{imn}(\bar{\psi}^i\Gamma^t{\psi}^i)(\bar{\psi}^j\Gamma^t{\psi}^m)(\bar{\psi}^k\Gamma^t{\psi}^n)
\eea
with the interaction vertices $\Gamma^\pm=\mathbbm{1}_4\pm\gamma_5$ for right- and left-handed channels, respectively. Here, one should note the Einstein summation convention for the flavor indices $i,j,k,m,n$ and the correspondences between $1,2,3$ and $u,d,s$. 

To our main concerns, we choose the following scalar and charged pseudoscalar condensates to be nonzero: $$\sigma_{\rm f}=\langle\bar\psi_{\rm f}\psi_{\rm f}\rangle,\ \Delta_\pi=\langle\bar{u}i\gamma^5d\rangle,\ \Delta_\pi^*=\langle\bar{d}i\gamma^5u\rangle.$$ For brevity, we set $\Delta_\pi=\Delta_\pi^*$ without loss of generality in the following. To facilitate the study, we'd like first to reduce ${\cal L}_{\rm tH}$ to an effective form with four-fermion interactions at most. By applying the Hartree approximation to contract a pair of quark and antiquark in each six-fermion interaction term~\cite{Klevansky:1992qe}, we immediately find
\begin{eqnarray}
{\cal L}_{\rm tH}^4
\!&=&\!-{K}\!\left\{\epsilon_{ijk}\epsilon_{imn}\sigma_i\!\left(\bar{\psi}^j{\psi}^m\bar{\psi}^k{\psi}^n
\!-\!\bar{\psi}^ji\gamma^5{\psi}^m\bar{\psi}^ki\gamma^5{\psi}^n\right)\!+\right.\nonumber\\
&&\left.\!\!\!\!2\Delta_\pi\!\!\left[\bar{s}{s}\!\left(\bar{u}i\gamma^5d\!+\!\bar{d}i\gamma^5u\!-\!\Delta_\pi\right)\!+\!\bar{s}i\gamma^5{s}\left(\bar{u}d\!+\!\bar{d}u\right)\right]\right\},
\end{eqnarray}
 where the second term in the brace is induced by $\pi^\pm$ condensations. Armed with the reduced Lagrangian density:
\begin{eqnarray}\label{LNJL4}
{\cal L}_{\rm NJL}\!&=&\!-V(L,L)\!+\!\bar\psi\!\left[i\slashed{\partial}\!-\!i\gamma^4\!\!\left(\!ig{\cal A}^4\!+\!Q_{\rm q}\mu_{\rm Q}\!+\!{\mu_{\rm B}\over3}\!\right)\!-\!m_0\right]\!\psi\nonumber\\
&&+G\sum_{a=0}^8\left[(\bar\psi\lambda^a\psi)^2+(\bar\psi i\gamma_5\lambda^a\psi)^2\right]+{\cal L}_{\rm tH}^4,
\end{eqnarray}
the left calculations can just follow the two-flavor case in principle. 

By contracting quark and antiquark pairs once more in the interaction terms of Eq.\eqref{LNJL4}, we find the quark bilinear form as
\begin{eqnarray}\label{LNJL2}
{\cal L}_{\rm NJL}^2\!\!=\!\bar\psi\left[i\slashed{\partial}\!-\! i\gamma^4\!\!\left(ig{\cal A}^4\!+\!Q_{\rm q}\mu_{\rm Q}\!+\!{\mu_{\rm B}\over3}\right)\!-\!m_{\rm i}\!-\!i\gamma^5\lambda^1\Pi \right]\psi,\nonumber\\
\end{eqnarray}
where the scalar and pseudoscalar masses are respectively
\begin{eqnarray}
m_i&=&m_{0i}-4G\sigma_i+2K(\sigma_j\sigma_k+\Delta_\pi^2\delta_{i3}),\nonumber\\
\Pi &=&(-4G+2K\sigma_3)\Delta_\pi\label{mpi}
\end{eqnarray}
with $i\neq j\neq k$. The $G$ and $K$ dependent terms in Eq.~\eqref{mpi} are from the $U_A(1)$ symmetric and anomalous interactions, respectively. According to Eq.\eqref{LNJL2}, $s$ quark decouples from $u,d$ quarks, so the gap equation for $\sigma_{\rm s}$ can be simply given by~\cite{Klevansky:1992qe}:
\begin{eqnarray}
\!\!\!\!\!\sigma_{\rm s}\!=\!\langle\bar{s}{s}\rangle\!=\!{\rm tr} \left[i\slashed{\partial}-i\gamma^4\!\!\left(ig{\cal A}^4\!+\!Q_{\rm q}\mu_{\rm Q}\!+\!{\mu_{\rm B}\over3}\right)\!-\!m_{\rm s} \right]^{\!-\!1}\!\!.\label{sigmas}
\end{eqnarray}
However, the $u$ and $d$ light quarks couple with each other through the non-diagonal pseudoscalar mass $\Pi$. Since $\mu_{\rm B}$ is usually small in early Universe, we can simply set $$m_{\rm 0u}=m_{\rm 0d}\equiv m_{\rm 0l},\ \sigma_{\rm u}=\sigma_{\rm d}\equiv\sigma_{\rm l}$$ in order to further carry out analytic derivations. Then, by following a similar procedure as the previous section, the explicit thermodynamic potential can be worked out for the bilinear terms as
\begin{widetext}
\bea
\!\!\!\!\!\!\Omega_{\rm bl}&=&\!-\!2N_c\!\!\int^\Lambda\!\!\!\!{\di^3k\over(2\pi)^3}\!\!\left[\sum_{t=\pm}\!E_{\rm l}^{\rm t}(k)\!+\!\epsilon_{\rm s}(k)\right]\!\!-\!2T\!\!\int\!\!\!{\di^3k\over(2\pi)^3}\!\!\sum_{\rm u=\pm}\!\!\left[\sum_{t=\pm}\!Fl\!\left(\!L,u,E_{\rm l}^{\rm t}(k),{\mu_{\rm Q}\!+\!2\mu_{\rm B}\over6}\!\right)\!\!+\!Fl\!\left(\!L,u,\epsilon_{\rm s}(k),{-\mu_{\rm Q}\!+\!\mu_{\rm B}\over3}\!\right)\!\right]
\eea
\end{widetext}
with the energy functions defined by
\bea
\!\!\!\!\epsilon_{\rm i}(k)\!=\!\sqrt{k^2\!+m_{\rm i}^2},\
E_{\rm l}^{\rm t}(k)\!=\!\sqrt{\left[\epsilon_{\rm l}(k)\!+t{\mu_{\rm Q}\over2}\right]^2\!+\Pi^2}.
\eea
Eventually, the coupled gap equations follow directly from the definitions of condensates:
\bea
&&\sigma_{\rm s}\equiv\langle\bar{s}{s}\rangle={\partial\Omega_{\rm bl}\over\partial m_{\rm s}},\
2\sigma_{\rm l}\equiv\langle\bar{u}{u}\rangle+\langle\bar{d}{d}\rangle={\partial\Omega_{\rm bl}\over\partial m_{\rm l}},\nonumber\\
&&2\Delta_\pi\equiv\langle\bar{u}i\gamma^5{d}\rangle+\langle\bar{d}i\gamma^5{u}\rangle={\partial\Omega_{\rm bl}\over\partial \Pi}\label{conds}
\eea
and the minimal condition $\partial_{\rm L}[V(L,L)+\Omega_{\rm bl}]=0$ as~\cite{Xia:2013caa}
\begin{widetext}
\bea
\sigma_{\rm s}
\!&=&\!-2N_c\int^\Lambda\!\!{\di^3k\over(2\pi)^3}{m_{\rm s}\over\epsilon_{\rm s}(k)}+2N_c\int\!\!{\di^3k\over(2\pi)^3}{m_{\rm s}\over\epsilon_{\rm s}(k)}\sum_{u=\pm}dV_1\left(\!L,u,\epsilon_{\rm s}(k),{-\mu_{\rm Q}\!+\!\mu_{\rm B}\over3}\!\right),\label{gaps}\\
2\sigma_{\rm l}\!&=&\!-2N_c\int^\Lambda\!\!{\di^3k\over(2\pi)^3}\sum_{t=\pm}{m_{\rm l}\over\epsilon_{\rm l}(k)}{\epsilon_{\rm l}(k)+t{\mu_{\rm Q}\over2}\over E_{\rm l}^{\rm t}(k)}+2N_c\int\!\!{\di^3k\over(2\pi)^3}\sum_{t,u=\pm}{m_{\rm l}\over\epsilon_{\rm l}(k)}{\epsilon_{\rm l}(k)+t{\mu_{\rm Q}\over2}\over E_{\rm l}^{\rm t}(k)}dV_1\left(\!L,u,E_{\rm l}^{\rm t}(k),{\mu_{\rm Q}\!+\!2\mu_{\rm B}\over6}\!\right),\label{gapl}\\
2\Delta_\pi\!&=&\!-2N_c\int^\Lambda\!\!{\di^3k\over(2\pi)^3}\sum_{t=\pm}{\Pi\over E_{\rm l}^{\rm t}(k)}+2N_c\int\!\!{\di^3k\over(2\pi)^3}\sum_{t,u=\pm}{\Pi\over E_{\rm l}^{\rm t}(k)}dV_1\left(\!L,u,E_{\rm l}^{\rm t}(k),{\mu_{\rm Q}\!+\!2\mu_{\rm B}\over6}\!\right),\label{gappi}
\eea
\bea
&&\!\!\!\!\!\!\!\!\!\!\!\!\!\!T^4\left[-\left(3.51\!-{2.47\over\tilde{T}}\!+\!{15.2\over\tilde{T}^{2}}\right)L\!+\!{1.75\over\tilde{T}^{3}}{12L(1-L)^2\over1\!-\!6L^2\!+\!8L^3\!-\!3L^4}\right]=6T\!\int\!\!{\di^3k\over(2\pi)^3}\!\sum_{u=\pm} \left[\sum_{t=\pm}dV_2\left(L,u,E^{\rm t}(k),{\mu_{\rm Q}\!+\!2\mu_{\rm B}\over6}\right)\right.\nonumber\\
&&\left.+dV_2\left(\!L,u,\epsilon_{\rm s}(k),{-\mu_{\rm Q}\!+\!\mu_{\rm B}\over3}\!\right)\right].
\eea
\end{widetext}
Note that $\Delta_\pi=0$ is a trivial solution of Eq.\eqref{gappi}, so $\Delta_\pi$ or $\Pi$ is still a true order parameter for $I_3$ isospin symmetry~\cite{Son:2000xc} in three-flavor case. The total self-consistent thermodynamic potential can be found to be
\bea
\Omega_{\rm 3f}&=&V(L,L)+\Omega_{\rm bl}+2G(\sigma_{\rm s}^2+2\sigma_{\rm l}^2+2\Delta_\pi^2)\nonumber\\
&&-4K(\sigma_{\rm l}^2+\Delta_\pi^2)\sigma_{\rm s}
\eea
by utilizing the definitions of condensates and their relations to scalar and pseudoscalar masses, refer to Eqs.\eqref{conds} and \eqref{mpi}. And the entropy, electric charge number and baryon number densities can be given according to the thermodynamic relations as
\begin{widetext}
\bea
\!\!\!\!\!\!s_{\rm 3f}&=&2\!\!\int\!\!{\di^3k\over(2\pi)^3}\!\sum_{t,u=\pm}\left[Fl\left(\!L,u,E^{\rm t}(k),{\mu_{\rm Q}\!+\!2\mu_{\rm B}\over6}\!\right)\!+\!{3\left(\!E^{\rm t}(k)\!-\!u\,{\mu_{\rm Q}\!+\!2\mu_{\rm B}\over6}\!\right)\over T}dV_1\left(\!L,u,E^{\rm t}(k),{\mu_{\rm Q}\!+\!2\mu_{\rm B}\over6}\!\right)\right]\nonumber\\
&&+2\!\!\int\!\!{\di^3k\over(2\pi)^3}\!\sum_{u=\pm}\left[Fl\left(\!L,u,\epsilon_{\rm s}(k),{-\mu_{\rm Q}\!+\!\mu_{\rm B}\over3}\!\right)\!+\!{3\left(\!E^{\rm t}(k)\!-\!u\,{-\mu_{\rm Q}\!+\!\mu_{\rm B}\over3}\!\right)\over T}dV_1\left(\!L,u,\epsilon_{\rm s}(k),{-\mu_{\rm Q}\!+\!\mu_{\rm B}\over3}\!\right)\right]\nonumber\\
&&+T^3\left\{{1\over2}\left(4\times3.51-3\times{2.47\over\tilde{T}}+2\times{15.2\over\tilde{T}^{2}}\right)L^2+{1.75\over\tilde{T}^{3}}\ln\left[1-6L^2+8L^3-3L^4\right]\right\},\label{s3f}\\
\!\!\!\!\!\!n_{\rm Q}^{\rm3f}&=&N_c\!\!\int^\Lambda\!\!{\di^3k\over(2\pi)^3}\!\sum_{t=\pm}t{\epsilon(k)\!+\!t{\mu_{\rm Q}\over2}\over E^{\rm t}(k)}-\!3\!\!\int\!\!{\di^3k\over(2\pi)^3}\sum_{t,u=\pm}t{\epsilon(k)\!+\!t{\mu_{\rm Q}\over2}\over E^{\rm t}(k)}dV_1\left(\!L,u,E^{\rm t}(k),{\mu_{\rm Q}\!+\!2\mu_{\rm B}\over6}\!\right)\nonumber\\
&&+\int\!\!{\di^3k\over(2\pi)^3}\!\sum_{t,u=\pm}u\,dV_1\left(\!L,u,E^{\rm t}(k),{\mu_{\rm Q}\!+\!2\mu_{\rm B}\over6}\!\right)-2\int{\di^3k\over(2\pi)^3}\!\sum_{t,u=\pm}u\,dV_1\left(\!L,u,\epsilon_{\rm s}(k),{-\mu_{\rm Q}\!+\!\mu_{\rm B}\over3}\!\right),\label{nQ3f}\\
\!\!\!\!\!\!n_{\rm B}^{\rm3f}&=&2\int{\di^3k\over(2\pi)^3}\!\sum_{t,u=\pm}u\,dV_1\left(\!L,u,E^{\rm t}(k),{\mu_{\rm Q}\!+\!2\mu_{\rm B}\over6}\!\right)+2\int{\di^3k\over(2\pi)^3}\!\sum_{t,u=\pm}u\,dV_1\left(\!L,u,\epsilon_{\rm s}(k),{-\mu_{\rm Q}\!+\!\mu_{\rm B}\over3}\!\right).\label{nB3f}
\eea
\end{widetext}

\subsection{The QEWD sector: free gases}\label{QEWD}
In free gas approximation, the thermodynamic potentials for the QEWD sector can be easily given by~\cite{Kapusta2006} 
\bea
\Omega_\gamma&=&2T\int{\di^3k\over(2\pi)^3}\log\left(1-e^{- k/T}\right),\\
\Omega_{\rm l}&=&-T\!\!\sum^{\rm i=e,\mu,\tau}_{u=\pm}\!\int\!\!{\di^3k\over(2\pi)^3}\!\left\{2\log\left[1+e^{- \left(\epsilon_{\rm i}(k)-u\,(-\mu_{\rm Q}+\mu_{\rm i})\right)/T}\right]\right.\nonumber\\
&&\left.+\log\left[1+e^{- (k-u\,\mu_{\rm i})/T}\right]\right\},
\eea
where the degeneracy is one for neutrinos and anti-neutrinos due to their definite chiralities. Then, the corresponding entropy, electric charge number and lepton flavor number densities can be derived directly as
\bea
s_\gamma&=&2\int{\di^3k\over(2\pi)^3}\left[-\log\left(1-e^{- k/T}\right)+{k/T\over e^{k/T}-1}\right],\\
s_{\rm l}&=&\sum^{\rm i=e,\mu,\tau}_{u=\pm}\int{\di^3k\over(2\pi)^3}\left\{2\log\left[1+e^{- \left(\epsilon_{\rm i}(k)-u\,(-\mu_{\rm Q}+\mu_{\rm i})\right)/T}\right]\right.\nonumber\\
&&+\log\left[1+e^{- (k-u\,\mu_{\rm i})/T}\right]\!+\!{2\left(\epsilon_{\rm i}(k)\!-\!u\,(\!-\mu_{\rm Q}\!+\!\mu_{\rm i})\right)/T\over1+e^{\left(\epsilon_{\rm i}(k)-u\,(-\mu_{\rm Q}+\mu_{\rm i})\right)/T}}\nonumber\\
&&\left.+{(k-u\,\mu_{\rm i})/T\over1+e^{(k-u\,\mu_{\rm i})/T}}\right\},
\eea
\bea
n_{\rm Q}^{\rm l}&=&2T\sum^{\rm i=e,\mu,\tau}_{u=\pm}\int{\di^3k\over(2\pi)^3}{-u\over1+e^{\left(\epsilon_{\rm i}(k)-u\,(-\mu_{\rm Q}+\mu_{\rm i})\right)/T}},\\
n_{\rm i}&=&-{\partial \Omega_{\rm l}\over \partial \mu_{\rm i}}=T\sum_{u=\pm}\int{\di^3k\over(2\pi)^3}\left[{2u\over1+e^{\left(\epsilon_{\rm i}(k)-u\,(-\mu_{\rm Q}+\mu_{\rm i})\right)/T}}\right.\nonumber\\
&&\qquad\qquad\ \ \ \left.+{u\over1+e^{(k-u\,\mu_{\rm i})/T}}\right],\ i=e,\mu,\tau.
\eea

Now, collecting contributions from both QEWD and QCD sectors, the total entropy, electric charge number and lepton number densities are respectively
\bea
\!\!\!\!\!\!s=s_\gamma\!+\!s_{\rm l}\!+\!s_{\rm 2f/3f},\,
n_{\rm Q}=n_{\rm Q}^{\rm l}\!+\!n_{\rm Q}^{\rm 2f/3f},\,
n_{\rm l}=\!\sum_{\rm i=e,\mu,\tau} \!\! n_{\rm i}
\eea
in the QCD epoch. To better catch the expansion nature of early Universe, we define several reduced quantities:
\bea
b=n_{\rm B}^{\rm2f/3f}/s,\ l=n_{\rm l}/s,\ l_{\rm i}=n_{\rm i}/s
\eea
by following the conventions. According to the introduction, $l$ is not so well constrained as $b$ from the observations, but the standard picture well predicts that $l=-51/28\, b$~\cite{Kolb:1983ni}. Furthermore, due to neutrino oscillations at late stage of the Universe, the lepton flavor densities $l_{\rm i}$ are not well constrained at the QCD epoch, thus we will take $l_{\rm e}$ and $l_{\rm \mu}$ as free variables in the following.

\section{Numerical results}\label{numerical}
To carry out numerical calculations, we get the muon mass from the Particle Data Group as $m_\mu=113\,{\rm MeV}$, simply set the electron mass $m_{\rm e}=0$, and suppress the contribution of heavy $\tau$ leptons for the QEWD sector. The model parameters are fixed for the QCD sector as the following~\cite{Zhuang:1994dw,Rehberg:1995kh}
\bea
\!\!\!\!\!\!\!\!&{\rm PNJL_{2f}}\!:& \ m_{\rm 0l}\!=\!5\,{\rm MeV},\, \Lambda\!=\!653\,{\rm MeV},\, G\Lambda^2\!=\!2.10;\\ 
\!\!\!\!\!\!\!\!&{\rm PNJL_{3f}}\!:&\  m_{\rm 0l}\!=\!5.5\,{\rm MeV},\, m_{\rm 0s}\!=\!140.7\,{\rm MeV},\nonumber\\ 
\!\!\!\!\!\!\!\!&&\  \Lambda\!=\!602.3\,{\rm MeV},\, G\Lambda^2\!=\!1.835,\, K\Lambda^5\!=\!12.36.
\eea

\begin{figure}[!htb]
	\begin{center}
		\includegraphics[width=8cm]{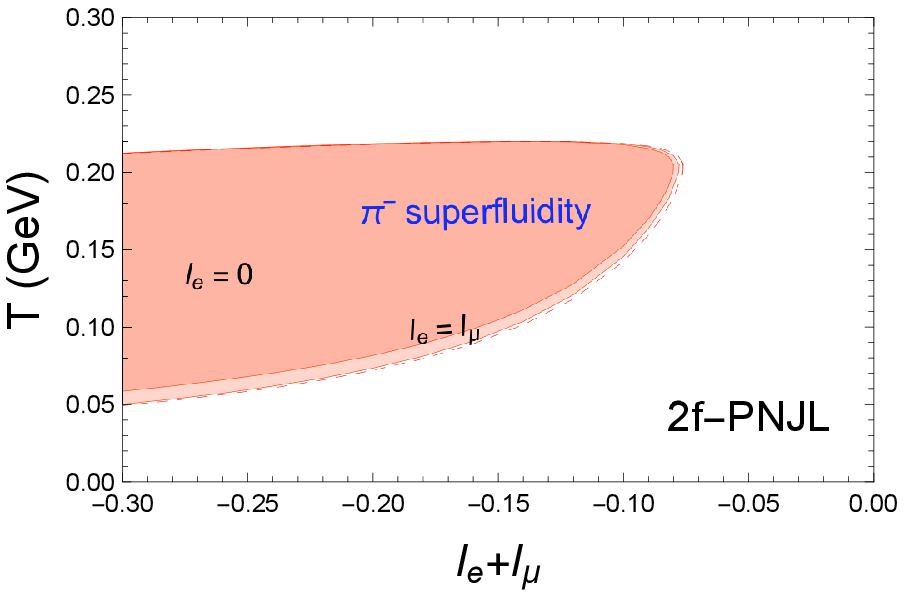}
		\includegraphics[width=8cm]{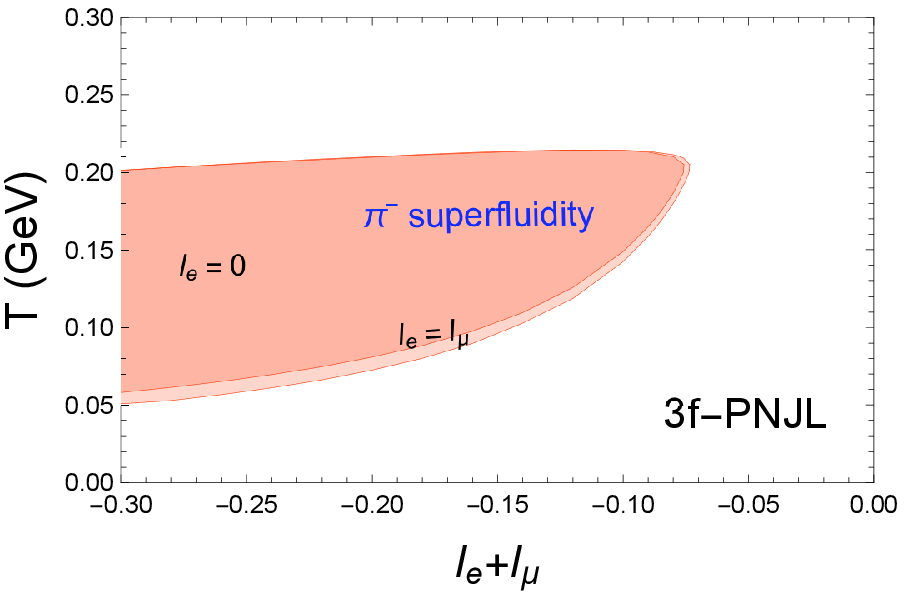}
		\caption{The $T-(l_{\rm e}+l_\mu)$ phase diagrams in two- (upper panel) and three-flavor (lower panel) PNJL models. The heavy and light shadows correspond to the pion superfluidity phase for $l_{\rm e}=0$ and $l_{\rm e}=l_\mu$ respectively with the total lepton asymmetry $l=-0.012$. The dashed line in the upper panel is the boundary for the total lepton asymmetry $l=-51/28 b$ and $l_{\rm e}=l_\mu$.}\label{Tl_2f}
	\end{center}
\end{figure}

The $T-(l_{\rm e}+l_\mu)$ phase diagrams of two-and three-flavor PNJL models are illuminated in Fig.\ref{Tl_2f}. As we can see, the ratio $l_{\rm e}/l_\mu$ and the effect of strange quarks are all not so important for determining the phase boundaries of pion superfluidity, especially the upper ones. And the calculations with the standard lepton asymmetry in the upper panel indicate that the uncertainty of $l$ plays a negligible role in the exploration of phase boundary. Compared to the threshold lepton flavor asymmetry $|l_{\rm e}+l_\mu|\sim0.1$ in the extrapolated lattice QCD calculations~\cite{Vovchenko:2020crk}, the values are consistently $|l_{\rm e}+l_\mu|\sim0.09$ at their top temperature $T_{\rm pc}=0.16\,{\rm GeV}$ in our evaluations. So recalling the effectivenesses of PNJL model and their criterion for the phase boundary, the agreement is remarkable! In advance, we obtain the upper boundaries of pion superfluidity to be consistently $T\sim0.21\,{\rm GeV}$ which is much larger than $T_{\rm pc}$, a well-known drawback of PNJL model~\cite{Xiong:2009zz}. Nevertheless, the upper boundaries are almost the corresponding pseudo-critical temperatures at zero chemical potentials, which then supports the setting of the upper boundary around $T_{\rm pc}$ in Ref.~\cite{Vovchenko:2020crk}. By the way, here the threshold lepton flavor asymmetries are $0.08$ and $0.075$ in the two- and three-flavor cases, respectively.

\begin{figure}[!htb]
	\begin{center}
		\includegraphics[width=8cm]{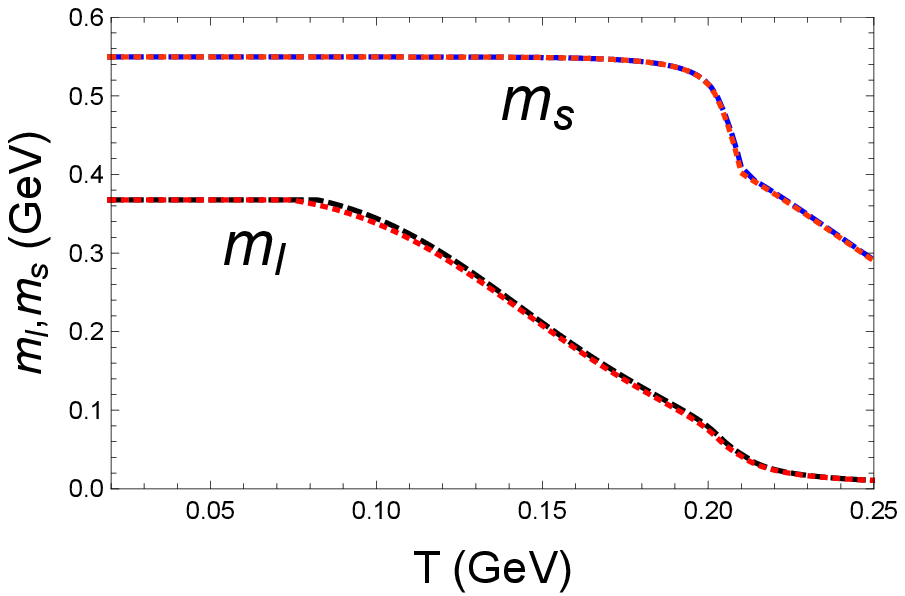}
		\includegraphics[width=8cm]{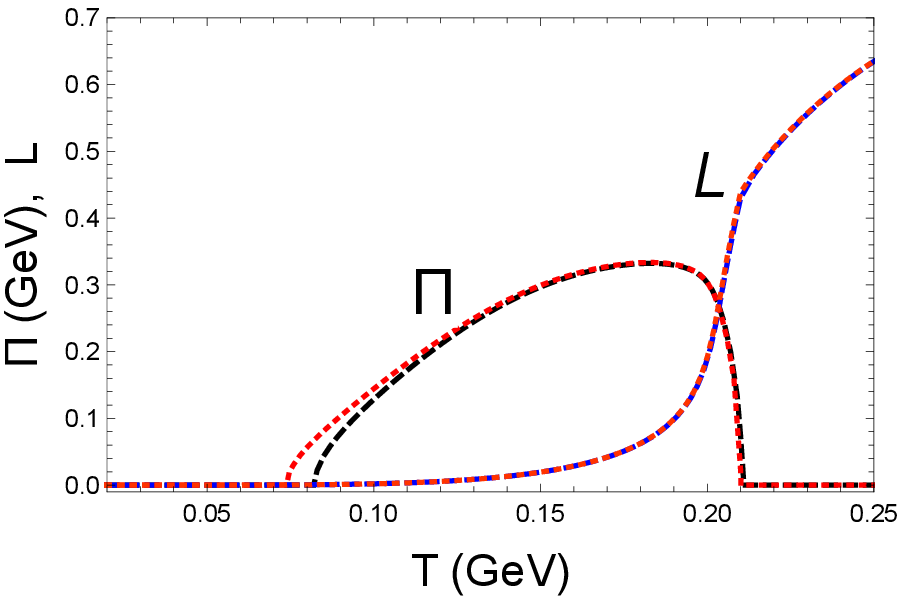}
		\caption{The order parameters $m_{\rm l}, m_{\rm s}$ (upper panel) and $\Pi, L$ (lower panel) as functions of temperature $T$ for $l=-0.012$ and $l_{\rm e}+l_\mu=-0.2$ in three-flavor PNJL model. The dashed and dotted lines correspond to the cases with $l_{\rm e}=0$ and $l_{\rm e}=l_\mu$, respectively.}\label{OP}
	\end{center}
\end{figure}

\begin{figure}[!htb]
	\begin{center}
		\includegraphics[width=8cm]{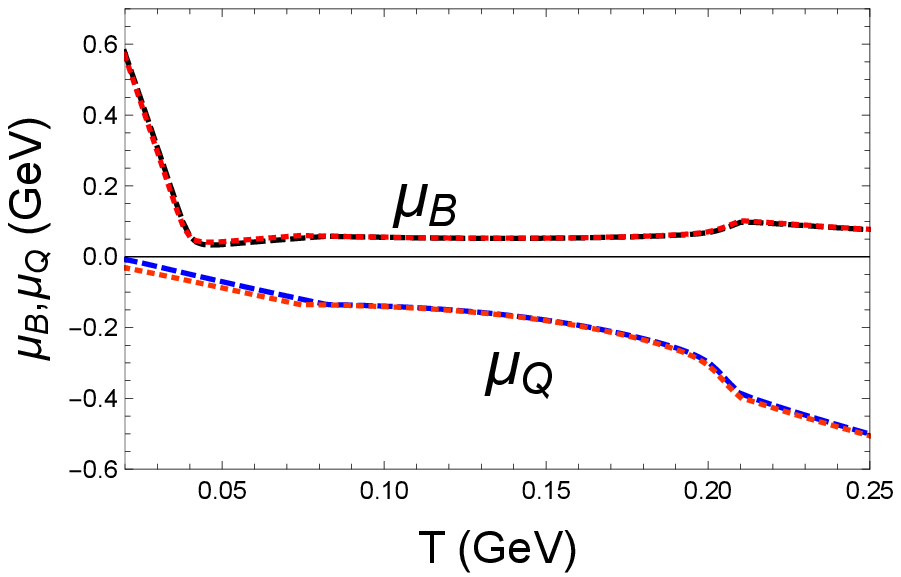}
		\includegraphics[width=8cm]{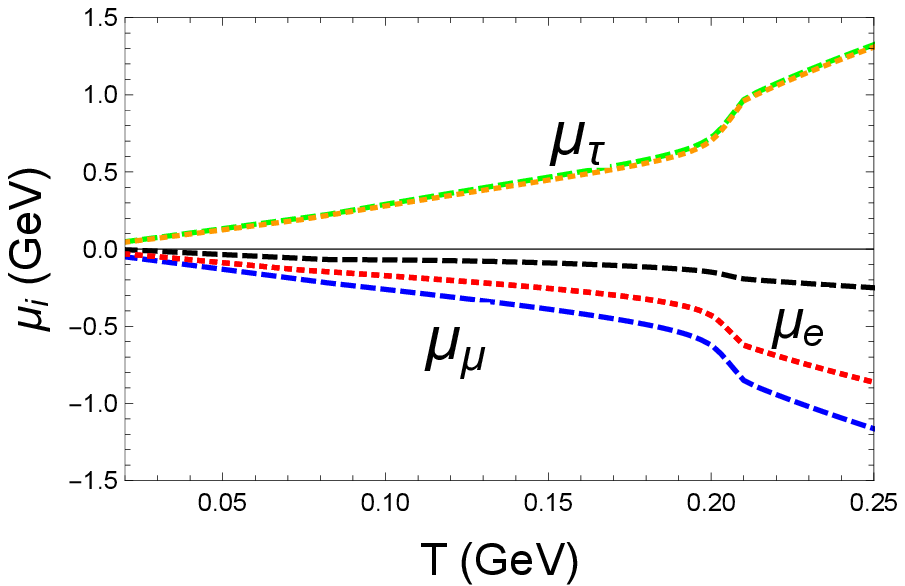}
		\caption{The cosmic trajectories of baryon and electric charge chemical potentials $\mu_{\rm B}, \mu_{\rm Q}$ (upper panel) and lepton flavor chemical potentials $\mu_{\rm e}, \mu_\mu,\mu_\tau$ (lower panel) as functions of temperature $T$ for $l=-0.012$ and $l_{\rm e}+l_\mu=-0.2$ in three-flavor PNJL model. The dashed and dotted lines correspond to the cases with $l_{\rm e}=0$ and $l_{\rm e}=l_\mu$, respectively. In the lower panel, the red dotted line denotes the overlapping of $\mu_{\rm e}$ and $\mu_{\mu}$.}\label{CT}
	\end{center}
\end{figure}

Now, we take the more realistic three-flavor PNJL model for example to show the features of cosmic trajectories at $l_{\rm e}+l_\mu=-0.2$, where the early Universe could evolve through the pion superfluidity phase. We compare two cases: $l_{\rm e}=0$ and $l_{\rm e}=l_\mu$, and demonstrate the order parameters and chemical potentials in Fig.\ref{OP} and Fig.\ref{CT}, respectively. As we can see, the effect of the ratio $l_{\rm e}/l_\mu$ is only important on the cosmic trajectories of the directly related quantities: $\mu_{\rm e}$ and $\mu_{\mu}$, and the results almost overlap with each other for other quantities. According to the lower panel of Fig.\ref{OP}, the pion condensate $\Pi$ shows a reentrant feature with $T$: although the decreasing at higher temperature can be easily understood as isospin symmetry restoration, the increasing at lower temperature is not trivial. Actually, the latter is due to the enhancement of $|\mu_{\rm Q}|$ with $T$ under the constraint of electric neutrality, see the upper panel in Fig.\ref{CT}. In Fig.\ref{OP}, it is also interesting to notice the consistency between the critical temperatures of $I_3$ isospin symmetry restoration and deconfinement transition at finite charge chemical potential~\cite{Xiong:2009zz}. 

Following the monotonous feature of $|\mu_{\rm Q}|$ in the upper panel of Fig.\ref{CT}, we find the criterion $|\mu_{\rm Q}|=m_\pi^{\rm v}$ to be well satisfied at the lower critical temperature $T_{\rm l}$; but that is no longer useful for the exploration of upper critical temperature $T_{\rm u}$, since the effective $\pi^\pm$ mass increases with $T$. Since the $|\mu_{\rm Q}|$ varies from $m_\pi^{\rm v}$ to $\sim3\,m_\pi^{\rm v}$ within the pion superfluidity phase, we can well recognize the BCS-BEC crossover therein~\cite{Sun:2007fc} as the early Universe cooled down. As expected, the baryon chemical potential is small except for very low temperature, which justifies our assumptions in the  QCD sector: $m_{\rm u}=m_{\rm d}$ and $L=L^*$. Furthermore, the opposite signs between $\mu_{\rm B}$ and $\mu_{\rm Q}$  and the abrupt jump of $\mu_{\rm B}$ at low temperature both qualitatively fit the findings in the extrapolated lattice QCD study~\cite{Wygas:2018otj}. We note that the exist of pion superfluidity at $T_{\rm l}$ only leaves visible sign in the behavior of $\mu_{\rm Q}$ but the entrance at $T_{\rm u}$ gives rise to important signs in all the chemical potentials. Compared to the signs ${\cal S}(\mu_{\rm e})= {\cal S}(\mu_\mu)={\cal S}(\mu_\tau)$ for the choice $l_{\rm e}=l_{\mu}=l_\tau$ in Ref.~\cite{Wygas:2018otj}, we find ${\cal S}(\mu_{\rm e})= {\cal S}(\mu_\mu)=-{\cal S}(\mu_\tau)$ due to the choice $|l_{\rm e}+l_{\mu}|\gg|l|$.

\section{Summary}\label{summary}
In this work, the possibility of pion superfluidity and the corresponding cosmic trajectories are self-consistently explored by varying the lepton flavor asymmetries within the PNJL model. The effects of strange quarks, total lepton asymmetry and the ratio $l_{\rm e}/l_\mu$ are all found to be mild on the phase boundary of pion superfluidity. Following the previous study in Ref.~\cite{Vovchenko:2020crk}, the phase boundary is constrained from both lower and upper sides in our study. At $T_{\rm cp}$, the lepton flavor asymmetry $|l_{\rm e}+l_\mu|$ is $0.09$ in our work, quite consistent with the threshold value $0.1$ obtained in Ref.~\cite{Vovchenko:2020crk}. However, with the pseudocritical temperature in PNJL model much larger than that from lattice QCD, we find that the threshold values shift to $0.08$ and $0.075$ for two- and three-flavor cases, respectively. 

According to the three-flavor example, the pion condensation shows a non-monotonous or reentrant feature as should be for the existences of both upper and lower second-order phase boundaries. While the sign of lower critical temperature is only visible in $\mu_{\rm Q}$, the signs of upper one show up in all the chemical potentials. So in principle, the phase transitions to and from pion superfluidity during the evolution of early Universe can be identified through the non-analytic features in the cosmic trajectories. Moreover, the critical $|\mu_{\rm Q}|$ at $T_{\rm u}$ is found to be around $3\,m_\pi^{\rm v}$, which is so large that it explains why the extrapolated lattice QCD study was unable to fix $T_{\rm u}$ at all~\cite{Vovchenko:2020crk}.

In Ref.\cite{Vovchenko:2020crk}, the equation of state of QCD+QEWD matter with different $|l_{\rm e}+l_\mu|$ was adopted to study the relic density of primordial gravitational wave; then inversely the observations of GW would help to constrain $|l_{\rm e}+l_\mu|$ and thus indirectly indicate whether pion superfluidity had happened or not in the QCD epoch. For a second order phase transition, we don't expect it to leave direct relics on GW, such as the pion superfluidity in this work; but for a first order one, the transition itself will give rise to a specific GW spectrum~\cite{Coleman:1977py,Callan:1977pt,Witten:1984rs}. Actually, with strong magnetic field presented in early Universe~\cite{Vachaspati:1991nm,Son:1998my,Grasso:2000wj}, the transitions relevant to pion superfluid, which is also superconductor, could be shifted from second order to first order. We will explore such interesting situation in more detail in our coming work. Hopefully, the advanced GW detectors, such as LIGO, SKA, LISA, and Tianqin, could help to capture the signals in future.

\emph{Acknowledgments}
G.C. is supported by the National Natural Science Foundation of China with Grant No. 11805290. L.H. is supported by the National Natural Science Foundation of China with Grant Nos. 11775123 and 11890712. P.Z. is supported by the National Natural Science Foundation of China with Grant No. 11975320.

\end{document}